\input epsf.tex

\def\a{\alpha}\def\b{\beta}\def\c{\chi}\def\d{\delta}\def\e{\epsilon}
\def\f{\phi}\def\g{\gamma}
\def\k{\kappa}\def\l{\lambda}\def\m{\mu}\def\n{\nu}\def
\p{\pi}\def\r{\rho}\def\s{\sigma}
\def\y{\eta}\def\x{\xi}
\def\ee{\varepsilon}

\def\de{\partial}\def\na{\nabla}
\def\inf{\infty}\def\id{\equiv}\def\mo{{-1}}\def\ha{{1\over 2}}
\def\app{\approx}

\def\({\left(}\def\){\right)}\def\[{\left[}\def\]{\right]}

\def\ex{{\rm e}}

\def\mn{{\mu\nu}}

\def\fe{field equations }\def\bh{black hole }\def\as{asymptotically }
\def\coo{coordinates }
\def\bg{background }

\def\ssy{spherically symmetric }
\def\cp{critical points }

\def\sch{Schwarzschild }\def\ads{anti-de Sitter }

\def\ab{asymptotic behavior }
\def\ie{i.e.\ }

\def\section#1{\bigskip\noindent{\bf#1}\smallskip}
\def\subsect#1{\bigskip\noindent{\it#1}\smallskip}
\def\nota{\footnote{$^\dagger$}}

\def\PRL#1{Phys.\ Rev.\ Lett.\ {\bf#1}}
\def\PR#1{Phys.\ Rev.\ {\bf#1}}\def\CQG#1{Class.\ Quantum Grav.\ {\bf#1}}
\def\NP#1{Nucl.\ Phys.\ {\bf#1}}

\def\JHEP#1{JHEP\ {\bf#1}}

\def\ref#1{\medskip\everypar={\hangindent 2\parindent}#1}
\def\beginref{\begingroup
\bigskip
\centerline{\bf References}
\nobreak\noindent}
\def\endref{\par\endgroup}

\def\cR{{\cal R}}\def\xx{|\x-\x_0|}\def\ss{\scriptstyle}
\font\small = cmr7

\magnification=1200

{\nopagenumbers
\line{}
\vskip40pt
\centerline{\bf Black brane solutions of Einstein-Maxwell-scalar}

\centerline{\bf theory with  Liouville potential}
\vskip40pt
\centerline{{\bf S. Mignemi}\nota{e-mail: smignemi@unica.it}}
\vskip10pt
\centerline {Dipartimento di Matematica, Universit\`a di Cagliari}
\centerline{viale Merello 92, 09123 Cagliari, Italy}
\smallskip
\centerline{and INFN, Sezione di Cagliari}

\vskip60pt\centerline{\bf Abstract}
\medskip
{\noindent
We investigate the global properties of black brane solutions of a three-parameter
Einstein-Maxwell model nonminimally coupled to a scalar with exponential
potential. The black brane solutions of this model have recently been investigated
because of their relevance for holography and for the AdS/condensed matter
correspondence.
We classify all the possible regular solutions and show that they exist only for a
limited range of values of the parameters and that their asymptotic
behavior either breaks hyperscaling invariance or has the form of a domain wall.
We also write down some exact solutions in Schwarzschild coordinates. }
\vskip160pt

\vfil\eject}

\def\ee{\ex^{2\y}}\def\ec{\ex^{2\c}}\def\en{\ex^{2\n}}\def\ef{\ex^{2\f}}
\def\er{\ex^{2\r}}\def\eax{\ex^{2a\x}}\def\ebx{\ex^{2b\x}}
\def\ve{\varepsilon}
\def\bb{black brane }

\section{1. The model}

The Einstein-Maxwell model nonminimally coupled to a scalar field with Liouville
potential (EML) has been studied in several contexts [1-5].
Models of this kind arise for example
as low-energy limits of supergravity and superstring theories [6].
They are especially relevant in the context of holography [3-4], since they are dual to
models exhibiting a breaking of the conformal symmetry leading to hyperscaling
violations [7].

Because of the nonminimal coupling, the standard no-hair theorems [8] do not hold for
these models, and their \bh solutions have unusual asymptotics and support nontrivial
scalar fields.
In particular, the \ssy\bh solutions have been classified in ref.\ [1], where it was
shown that all such solutions have non-standard asymptotic behavior.
Because the potential for the scalar field does
not admit local extrema,  these models do not allow for AdS vacua.
Hence, also the Reissner-Nordstrom-AdS \bh solutions are forbidden.
It should be noticed that instead AdS$_{2}\times S^{2}$ solutions can exist.

Particularly interesting are also the solutions exhibiting planar symmetry,
especially black branes. From an analytical point of view, planar solutions have the
advantage of being easier to find in closed form than \ssy ones, since their calculation
requires one less integration, but are especially interesting  because of
their relevance for holography and the AdS/condensed matter correspondence [3,4].

Although EML models do not allow for AdS vacua, they still could admit asymptotic
solutions preserving scaling isometries. These solutions would be asymptotic to Lifshitz
spacetimes [9,10], i.e.\ spacetimes presenting a scale isometry
under which timelike and spacelike coordinates transform with
different exponents. This property may be relevant
for the holographic description of quantum phase transitions.

In the general case, the scaling isometry is broken and the metrics
only transform covariantly under scale transformations [5].
To this class belong in particular solutions for which the
Poincar\'e invariance  of the brane is preserved. In the literature they are often
referred to as domain wall solutions [11,4,12-14], in analogy with the domain wall
solutions of supergravity theories. In a EML context they were studied in [4].
For a wide range of values of the the parameters, domain wall solutions are conformal
to AdS and allow for an holographic interpretation [11,4].

It has been shown that the holographic interpretation of Einstein-Maxwell-scalar theories
gives rise  to a very rich phenomenology  in the dual QFT.
This includes phase transitions triggered by scalar condensates
and non-trivial transport properties of the dual field theory [2-5,10,12,15].
From the holographic point of view the difficulty connected with the absence of an AdS
vacuum, and hence of an ultraviolet fixed point in the dual QFT can easily be
circumvented.
In fact EML models can be considered as the near-horizon, near-extremal
description of an Einstein-Maxwell-dilaton gravity model whose scalar field potential
allows for a local maximum, and hence for an ultraviolet fixed point [16].

Very recently, it has been realized  that the solutions of the EML theory
represent the main example  of scale covariant models that lead to
hyperscaling violation in the dual field theory [7].
They are therefore a very promising
framework for the holographic description of hyperscaling violation
in condensed matter critical systems  (e.g.\ Ising models [17]).
Moreover they  have been used for the
description of Fermi surfaces and for the related area-law
violation of the entanglement entropy [7,18].

Although planar solutions of the EML model have been already investigated and several
partial results have been obtained [1-4], in this paper we systematize the previous
findings and obtain general results for all values of the parameters of the model.
In particular, we study the phase space of the dynamical system associated with the planar
solutions of the model in four dimensions, using the same approach adopted in ref.\ [1]
for the \ssy case, and classify all the possible black brane solutions admitting regular
horizon, according to their asymptotic behavior.
It turns out that all solutions have nonstandard asymptotics, of domain wall or
hyperscaling-violating form, but not of \ads or Lifshitz form. Moreover, only for a very
limited range of values of the parameters of the model regular \bb solutions are possible.

We also obtain some exact solutions for special values of the parameters.
Some of these solutions have already been obtained in [2], using a different
parametrization of the metric, that gives rise to complicated field equations, and
less clear interpretation.

Since the main purpose of this paper is the classification of the black brane solutions
of the EML theory with regular horizon and asymptotic regions, we shall not investigate
other possible solutions (namely solutions containing naked singularities or cosmological
horizons).
We also do not discuss the physical applications of the model, for example to holography,
except for some considerations on thermodynamics.
A thorough investigation of these topics can be found in [3-5].

\section{2. Action and field equations}
We consider the action
$$I={1\over16\p}\int\sqrt{-g}\ d^4x\,[R-2(\de\f)^2-\ex^{-2g\f}F^2+2\l\ex^{-2h\f}],\eqno(2.1)$$
where $F_\mn$ is a Maxwell field, $\f$ is a scalar field and $g$ and $h$ are two real
parameters.  Some special cases are well known: for $h\to\inf$ one obtains the GHS model [19,20],
for $h=0$ the GHS model with a cosmological constant. For $g\to\inf$ one gets a Liouville
model, for $g=0$ the minimally coupled Einstein-Maxwell theory with a Liouville potential. For
$h=-g=\pm1$, the action can be derived from string theory [11], for $h=-1/g=\pm\sqrt3$ from the
Kaluza-Klein reduction of a five-dimensional model.

The \fe read
$$\eqalign{&G_\mn=2\de_\m\f\de_\n\f-g_\mn\de^\r\f\de_\r\f+2\ex^{-2g\f}\left(F_{\m\r}F_\n^{\ \r}-
{1\over4}\,g_\mn F^2\right)+\l g_\mn\ex^{-2h\f},\cr
&\na^2\f=-{g\over2}\,\ex^{-2g\f}F^2+h\l\ex^{-2h\f},\cr
&\na^\m(e^{-2g\f} F_\mn)=0.}\eqno(2.2) $$

We look for electrically charged solutions with planar symmetry. Magnetic solutions can be
obtained by duality. More precisely, electrically charged solutions of the model with parameter
$g$ are magnetically charged solutions of the model with parameter $-g$ [2].

It is useful to parametrize the metric as\footnote{*}{This parametrization was originally
introduced in [20] and then used in several papers for discussing the phase space of solutions
of more complicated models [21,1]. We refer the reader to these papers for a detailed discussion
of these methods.}
$$ds^2=-\ex^{2\n} dt^2+\ex^{2\n+4\r}d\x^2+\ex^{2\r}(dx^2+dy^2),\eqno(2.3)$$
where $\n$, $\r$ and $\f$ are functions of the radial coordinate $\x$.

In these coordinates, the Maxwell equation is solved by
$$\qquad F_{t\x}=\ex^{2(\n+g\f)}Q,\eqno(2.4)$$
with $Q$ the electric charge,
and the remaining field equations can then be put in the form
$$\eqalignno{&\n''=\l\ee+Q^2\ec,&(2.5)\cr
&\r''=\l\ee-Q^2\ec,&(2.6)\cr
&\f''=h\l\ee+gQ^2\ec,&(2.7)\cr
&\r'^2+2\r'\n'-\f'^2-\l\ee+Q^2\ec=0,&(2.8)}$$
where
$$\y=\n+2\r-h\f,\qquad\c=\n+g\f.\eqno(2.9)$$
Writing the equations (2.5), (2.7) and (2.8) in terms of the independent variables
$\y$, $\c$ and $\r$, one obtains
$$\eqalignno{&\y''=(3-h^2)\l\ee-(1+gh)Q^2\ec,&(2.10)\cr
&\c''=(1+gh)\l\ee+(1+g^2)Q^2\ec,&(2.11)\cr
&\a\r'^2+\c'^2+\y'^2-2\y'\c'-2\g_1\r'\y'+2\g_2\r'\c'+(g+h)^2(\l\ee-Q^2\ec)=0,&(2.12)}$$
where
$$\a=4+3g^2+2gh-h^2,\qquad\g_1=2+g^2+gh,\qquad\g_2=2-h^2-gh.\eqno(2.13)$$
The case $\a=0$ is singular, since (2.10) and (2.11) are no longer independent.

In the following discussion, it will be important to know in what range of values
of $g$ and $h$ the constants $\a$, $\g_1$ and $\g_2$ are positive. This is shown
in fig.\ 1, and we shall implicitly assume this result in the rest of the paper.
In particular,
it is useful to observe that $\g_1<0$ implies that $\a$ and $\g_2$ are negative and
that if $h^2<3$, $\a$ and $\g_1$ are positive.

The system (2.5)-(2.8) is invariant under the shift
$$\n\to\n-g\ve,\qquad\r\to\r+{g+h\over2}\,\ve,\qquad\f\to\f+\ve,\eqno(2.14)$$
with constant parameter $\ve$. This invariance, which is not present in the \ssy
case, facilitates the solution of the system. In fact, it implies the existence of
a conserved quantity, that can be obtained combining (2.6), (2.10) and (2.11).
Indeed,
$$\r''={\g_1\y''-\g_2\c''\over\a},\eqno(2.15)$$
and hence
$$\g_1\y'-\g_2\c'-\a\r'=b,\eqno(2.16)$$
with $b$ an integration constant.
Substituting in (2.12) one gets
$$-{b^2\over(g+h)^2}+(1+g^2)\y'^2+2(1+gh)\y'\c'-(3-h^2)\c'^2-\a(\l\ee-Q^2\ec)=0.\eqno(2.17)$$
Eqs.\ (2.10) and (2.11) form a dynamical system for the variables $\y$ and $\c$, subject
to the constraint (2.17).

In view of the following discussion, it is useful to write down the functions $\n'$ and $\f'$
in terms of $\y'$ and $\c'$, taking into account (2.16):
$$\eqalignno{\n'=&{1\over\a}\[(4+gh-h^2)\c'+g(g-h)\y'+{2gb\over g+h}\],&\cr
\f'=&{1\over\a}\[(3g+h)\c'-(g-h)\y'-{2b\over g+h}\].&(2.18)}$$
\bigskip
\centerline{\epsfysize=8truecm\epsfbox{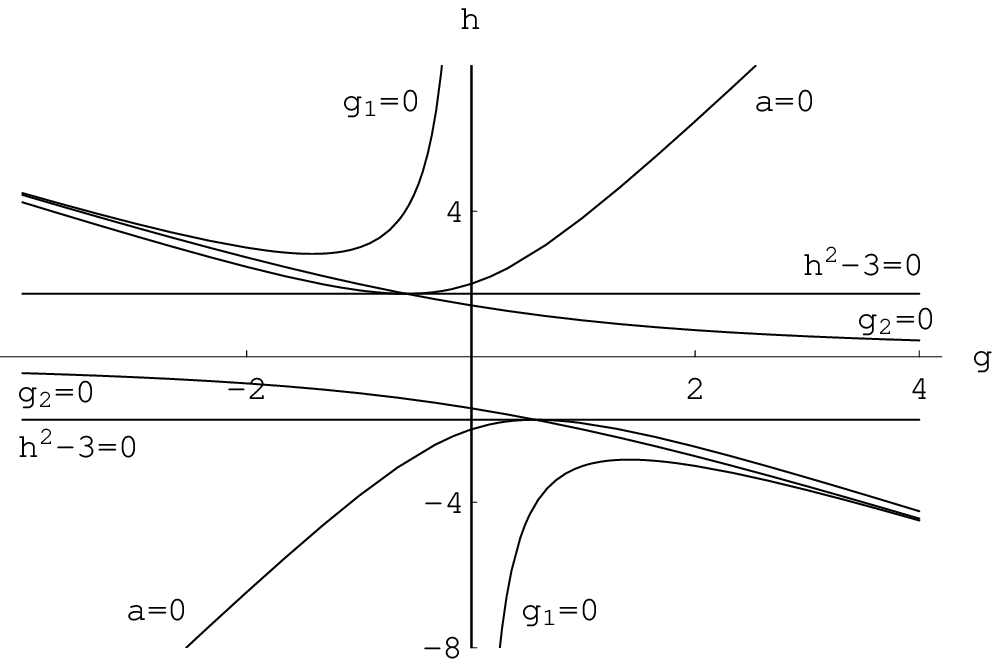}}
\medskip
{\baselineskip10pt
\centerline{\noindent\small Fig.\ 1: The curves $\ss\a\ =\ 0$, $\ss\g_1\ =\ 0$, $\ss\g_2\ =\ 0$
and $\ss 3\;-\;h^2\ =\ 0$ in the $\ss g-h$ plane. At the origin all functions are positive.}

\section{3. Exact solutions}
In a few special cases the \fe can be solved exactly. These solutions
are useful for the understanding of the general case.
\bigbreak
\subsect{A. Neutral solutions, $Q=0$}
A simple case in which an exact solution can be found is when the electric charge
vanishes, $Q=0$. This is a minimally coupled gravity-scalar model with exponential potential.
The standard no hair theorems [8] do not apply because the
solutions are not spherically symmetric, and hence nontrivial solutions are possible.

The \fe (2.5)-(2.8) reduce to
$$\eqalignno{&\n''=\r''={1\over h}\,\f''=\l\ee,&(3.1)\cr
&\r'^2+2\n'\r'-\f'^2-\l\ee=0.&(3.2)}$$
In addition to (2.14), the system now enjoys a further symmetry, for $\n\to\n+\k$,
$\r\to\r+h\,\k$, $\f\to\f+2\k$,
with constant parameter $\k$, that permits to completely integrate the system.

Assuming $h^2\ne3$, from (3.1) it follows that
$$\n'={\y'+b\over3-h^2},\qquad\r'={\y'+d\over3-h^2},\qquad\f'={h\y'+c\over3-h^2},\eqno(3.3)$$
with integration constants $b$, $d$, $c$, with $c=(b+2d)/h$.
Solving (2.10) with $Q=0$ yields
$$\y'^2=(3-h^2)\l\ee+a^2,\qquad\l\ee={4a^2\eax\over(3-h^2)(1-\eax)^2}.
\eqno(3.4)$$
Substituting (3.3) and (3.4) into (3.2), one gets a constraint between the parameters
of the solution,
$$h^2(3-h^2)a^2+2(h^2-2)bd+(h^2-4)d^2=b^2.\eqno(3.5)$$

For $\x\to0$, the radial function $\ex^\r\sim\ex^{(\y+d\x)/(3-h^2)}$ goes to infinity if
$h^2<3$, or to zero if $h^2>3$. In the first case, we identify this limit with spatial infinity.
The other relevant limit is for $\x\to\pm\inf$. If the solution possesses a regular horizon,
$e^\r$ must be a nonvanishing constant in that limit.
This request enforces the choice $d=-a$, and hence $b=(2-h^2)a$, $c=-ha$.

Integrating (3.3) and substituting the solution (3.4), one can obtain the explicit form
of the metric functions $\n$, $\r$ and $\f$. Rather than pursuing with the present
coordinates it is however convenient, in order to get a more transparent interpretation
of the solutions, to write the metric in a \sch form, namely,
$$ds^2=-Udt^2+U^\mo dr^2+R^2(dx^2+dy^2),\eqno(3.6)$$
where $U(r)=\ex^{2\n}$, $R(r)=\ex^\r$, and $F_{tr}={Q\over R^2}$.
In this gauge, the \fe take the form
$$\eqalign{{d^2R\over dr^2}&=-R\({d\f\over dr}\)^2,\cr
{d\over dr}\(UR{dR\over dr}\)&=-{Q^2\over R^2}\ex^{2g\f}+\l R^2\ex^{-2h\f},\cr
{d\over dr}\(UR^2{d\f\over dr}\)&=g{Q^2\over R^2}\ex^{2g\f}+h\l R^2\ex^{-2h\f},}\eqno(3.7)$$

The new coordinate $r$ is related to $\x$ by
$$r=\int\ex^{2(\n+\r)}d\x\app(1-\eax)^{-(1+h^2)/(3-h^2)}.\eqno(3.8)$$
where $\app$ means modulo a constant factor. It follows that
$$\eax\app1-\m r^{-(3-h^2)/(1+h^2)},\qquad\ee\app\left(1-\m r^{-(3-h^2)/(1+h^2)}\right)
r^{2(3-h^2)/(1+h^2)},\eqno(3.9)$$
with $\m$ a constant.
From the solutions obtained above, one can finally write down the metric functions in terms
of $r$ and the new integration constants $A$, $B$, $C$, $\m$, as
$$\eqalign{&R=\ex^\r=Ar^{1/(1+h^2)},\qquad U=\en=B\left(1-\m r^{-(3-h^2)/(1+h^2)}\right)
r^{2/(1+h^2)},\cr&\ef=C^2r^{2h/(1+h^2)}.}\eqno(3.10)$$
The constants can be fixed substituting (3.10) into (3.7). It turns out that $\m$ is a free
parameter,
while two of the constants, say $A$ and $C$, can be set to 1 by rescaling the coordinates.
Then
$$B={(1+h^2)^2\l\over3-h^2}.$$
These solutions have been found in [13] using a different method.

It is easy to verify that the scalar curvature is proportional to
$$\cR=3(h^2-2)r^{-2h^2/(1+h^2)}-\m h^2r^{-(3+h^2)/(1+h^2)},$$
and hence the only singularity is at $r=0$. Therefore, for positive $\l$ and $h^2<3$,
the solutions (3.10) describe a one-parameter family of black
brane solutions with horizon at $r_h=\m^{(1+h^2)/(3-h^2)}$ and domain-wall asymptotics,
that reduces to planar anti-de Sitter if $h=0$.
Solutions exist also for negative $\l$, if $h^2>3$. In this case the
mass term dominates and an asymptotic region exists only if $\m<0$, with
$U\sim r^{-(1-h^2)/(1+h^2)}$. However, a naked singularity occurs at $r=0$.
We conclude that solutions with a regular horizon and regular infinity exist
only if $\l>0$ and $h^2<3$.

\subsect{B. Planar GHS solutions, $\l=0$}
Another case in which it is possible to obtain exact solutions is when the potential
vanishes, $\l=0$. This is the planar extensions of the well-known GHS solutions [19-20].

Also in this case a new symmetry is present that enforces complete integrability.
The \fe (2.5)-(2.8) reduce to
$$\eqalignno{&\n''=-\r''={\f''\over g}=Q^2\ec,&(3.11)\cr
&\r'^2+2\n'\r'-\f'^2+Q^2\ec=0,&(3.12)}$$
and the system is invariant under (2.14) and under $\n\to\nu+\k$, $\r\to\r$, $\f\to\f-\k$.

From (3.11) follows that
$$\n'={\c'+b\over1+g^2},\qquad\r'={-\c'+d\over1+g^2},\qquad\f'={g\c'+c\over1+g^2},\eqno(3.13)$$
with $c=-b/g$, and eq.\ (2.11) is solved by
$$\c'^2=(1+g^2)Q^2\ec+a^2,\qquad Q^2\ec={4a^2\eax\over(1+g^2)(1-\eax)^2}.
\eqno(3.14)$$
Moreover, substituting in (3.12) one gets the condition
$$-(1+g^2)a^2+2bd+d^2={b^2\over g^2}.\eqno(3.15)$$

In this case, $\ex^\r\sim\ex^{-\c+d\x}$ and vanishes for $\x\to0$.
As in the previous case, a horizon is present if $\ex^\r$ goes to a constant at $\x\to\inf$,
\ie if $d=a$. It follows that $b=g^2a$, $c=-ga$.
Defining the coordinate $r$ as in (3.8), with a suitable choice of the integration constant
one obtains
$$r\app\ex^{2a\x}-1.\eqno(3.16)$$
Passing to \sch coordinates (3.6),
$$\eqalign{&R=\ex^\r=Ar^{1/(1+g^2)},\qquad
U=\en=B r^{-2/(1+g^2)}(1+\m r),\cr
&\ef=C^2r^{-2g/(1+g^2)}.}\eqno(3.17)$$
Also in this case, substituting in (3.7), one can easily check that $\m$ is free,
$A$ and $C$ can be set to 1, and
$$B={Q^2\over1+g^2}.$$

The scalar curvature is proportional to
$$\cR=r^{-2(2+g^2)/(1+g^2)}+\m r^{-(3+g^2)/(1+g^2)},$$
and hence diverges at $r=0$.
The properties of the solutions depend on the sign of the free parameter $\m$, which
dictates the asymptotic behavior. An asymptotic region is present only if $\m>0$, with
$U\sim r^{-(1-g^2)/(1+g^2)}$.
In this case, a horizon is present at $r=-\m$, but a naked singularity occurs at the origin,
so these are not regular \bb solutions.
For $\m<0$, the horizon is of cosmological type, but a naked singularity is still present.
Therefore, no regular \bb solutions exist in this case.

\subsect{C. Special case $\c'=\y'$}
Finally, an exact special solution can be found for generic values of the parameters of the
action, when $\c'=\y'$, i.e. $\c=\y+\log K$, with
constant $K$.
In this case,
$$\r'={(g+h)^2\y'+b\over\a},\quad\n'={(g+h)(4+g^2-h^2)\y'-2gb\over\a(g+h)},
\quad\f'=2\,{(g+h)^2\y'+b\over\a(g+h)},\eqno(3.18)$$
with $b$ an integration constant. Moreover, comparing (2.10) and (2.11), one gets
$$K^2={\g_2\over\g_1}\,{\l\over Q^2},\eqno(3.19)$$
and (2.12) becomes
$$-{b^2\over(g+h)^4}+\y'^2={\a\over\g_1}\,\l\ee.\eqno(3.20)$$
with $\a$, $\g_1$ and $\g_2$ given by (2.13).

Solving (2.10), with $Q^2\ec={\g_2\over\g_1}\l\ee$, one then obtains
$$\y'^2={\a\over\g_1}\,\l\ee+a^2,\qquad\l\ee={4a^2\g_1\eax\over\a(1-\eax)^2},
\eqno(3.21)$$
with $a$ an integration constant. Substituting this result in (3.20), one gets
$b^2=(g+h)^4a^2$. The radial function is given now by $\ex^\r=\ex^{[(g+h)^2\y+b\x]/\a}$,
and if $\a$ is positive diverges for $\x\to0$, signalling the presence of an asymptotic
region.
The request that $\ex^\r$ goes to a constant for $\x\to-\inf$, necessary to ensure the
presence of a horizon, enforces instead the choice $b=-(g+h)^2a$.

Defining then the radial coordinate $r$ as in (3.8), one obtains
$$r\app(1-\eax)^{-\b/\a},\eqno(3.22)$$
where
$$\b=4+(g+h)^2.\eqno(3.23)$$
Again, one can write the metric in the form (3.6). One has
$$\eqalign{&R=\ex^\r=Ar^{(g+h)^2/\b},\qquad U=\en=B\left(1-\m r^{-\a/\b}\right)
r^{2(4+g^2-h^2)/\b},\cr&\ef=C^2r^{4(g+h)/\b}.}\eqno(3.24)$$
The constants can be determined substituting in (3.7). In particular, one  of
them, say $C$, can be set to 1, while it is
easy to check that $\m$ is a free parameter. The remaining constants  result in
$$A^2=\sqrt{{\g_1\over\g_2}\,{Q^2\over\l}},\qquad B={\b^2\over\a\g_1}\,\l.$$
These solutions have been obtained in different \coo in [2].

Also in this case the scalar curvature is the sum of two terms, one proportional to
$r^{-4h(g+h)/\b}$, and the other to $\m r^{-(4+3(g+h)^2)/\b}$. While the second
term can diverge only at $r=0$, the first is singular either at $r=0$ or at $r=\inf$,
depending on the value of the parameters $g$ and $h$.

The properties of the solutions depend on the values of the parameters $g$, $h$ and
$\l$. It turns out that solutions which are regular at infinity and possess a regular
horizon are possible only if $\l$, $\a$, $\g_1$ and $\g_2$ are all positive and $h(g+h)>0$.
In that case the solutions (3.24) describe
charged black branes, with horizon at $r_h=\m^{\b/\a}$ and hyperscaling-violating
asymptotic behavior, that in the special case $g=2/h-h$ is enhanced to that
of a domain wall.

\section{4. Thermodynamics}
Before passing to the study of the global structure of the space of solutions of the
system (2.10)-(2.12), we calculate the thermodynamical parameters of the exact black brane
solutions found in the previous section. We omit the $\l=0$ case, since it always contains
naked singularities.

\subsect{A. $Q=0$.}
These solutions describe black branes if $\l>0$ and $h^2<3$.

In general, the mass density $m$ can be defined as [10]
$$m=\lim_{r\to\inf}-{1\over8\p}\sqrt U\[\sqrt U\ {dR^2\over dr}-\sqrt U\ {dR^2\over dr}\,
\bigg|_{\rm bg}\ \],\eqno(4.1)$$
where we have subtracted a term corresponding to the solution evaluated on the background
with $r_h=0$. For the solutions (3.10) this gives
$$m={\l\over8\p}\ {1+h^2\over3-h^2}\,\m.\eqno(4.2)$$

The temperature $T$ can be obtained calculating the periodicity of the Euclidean section,
as
$$T={1\over4\p}{dU(r_h)\over dr}.\eqno(4.3)$$
In our case
$$T={\l\over4\p}\ (1+h^2)\,\m^{(1-h^2)/(3-h^2)}.\eqno(4.4)$$

Finally, the entropy density $s$ is simply obtained as
$$s={1\over4}R^2(r_h).\eqno(4.5)$$
Hence,
$$s={1\over4}\m^{2/(3-h^2)}.\eqno(4.6)$$
It follows that $dm=Tds$ and $m={Ts\over3-h^2}$.

For $m\to0$, the entropy of the brane vanishes, while the temperature vanishes if $h^2<1$
and diverges if $h^2>1$. When $h^2=1$ the temperature is independent of the mass.

\subsect{B. $\c'=\y'$}
These solutions are valid if $\l,\a,\g_1,\g_2>0$ and $h(g+h)>0$.
Using the previous formulae, one obtains for the solutions (3.23)
$$m={(g+h)^2\over8\p}\
{\b\over\a}\sqrt{\l\,Q^2\over\g_1\g_2}\ \m,\eqno(4.7)$$
and
$$T={\l\over4\p}\ {\b\over\g_1}\,\m^{(4+g^2-2gh-3h^2)/\a},\eqno(4.8)$$

$$s={1\over4}\sqrt{{\g_1\over\g_2}\,{Q^2\over\l}}\ \m^{2(g+h)^2/\a}.\eqno(4.9)$$
It is easy to see that $dm=Tds$ and $m={2(g+h)^2\over\a}\,Ts$. Also in this case,
for $m\to0$ the entropy vanishes, while the temperature vanishes if $\a>2(g+h)^2$
or diverges otherwise.

\section{5. The dynamical system.}
For arbitrary values of the parameters $g$ and $h$, following the methods of
ref.\ [21,1], eqs.\ (2.10),(2.11) and (2.17)
can be put in the form of a dynamical system, by defining
$X=\c'$, $Y=\y'$, $P=Q\,\ex^\c$, $Z=\sqrt{|\l|}\,\ex^\y$, with
$$\eqalign{&X'=(1+gh)\e Z^2+(1+g^2)P^2,\cr&Y'=(3-h^2)\e Z^2-(1+gh)P^2,\cr &Z'=YZ.}
\eqno(5.1)$$
The independent variables are $X$, $Y$ and $Z$, while $P^2$ is defined as
$$P^2=\e Z^2-{(1+g^2)Y^2-(3-h^2)X^2+2(1+gh)XY\over\a}+{b^2\over(g+h)^2\a},\eqno(5.2)$$
where $\a$ is given by (2.13) and $\e=+1$ if $\l>0$ or $\e=-1$ if $\l<0$.
For each value of the free parameters $b$ and $Q$, we can now discuss the structure of
the phase space of the dynamical system. In the following discussion we shall not
consider the limit cases $Q=0$ and $\l=0$, since these have already been examined in
sect.\ 3. Their properties can however be useful in order to understand the general case.

The details of the phase space depend on the location of the critical points. All the
solutions of interest connect a critical point at finite distance with a critical point
at infinity. We shall therefore undertake the study of these points. Since
the system is invariant for $Z\to-Z$, it is sufficient to discuss the $Z>0$ portion
of phase space. Moreover, the system is invariant for $X\to-X$, $Y\to-Y$, together
with $\x\to-\x$, and hence to each critical point with $X>0$ corresponds a \cp with
$X<0$, with the direction of the trajectories reversed. Therefore, in the following we
list only the \cp with positive $X$.
\subsect{Critical points at finite distance}
For positive $X$, the critical points at finite distance are
attained in the limit $\x\to-\inf$. In the following discussion it will be important
to  study the behavior of the radial function $\ex^\r$ near the critical point, since
it indicates to what physical region these points correspond.
While in the most common cases [21] the critical points at finite distance correspond
to a finite value of $\ex^\r$ and hence to a region of spacetime at finite distance,
in the present case, for some values of the parameters, they may correspond to
$\ex^\r\to\inf$, and hence to spatial infinity.

The \cp at finite distance lie at $Z_0=P_0=0$, and hence their \coo $X_0$, $Y_0$
satisfy the equation
$$(1+g^2)Y_0^2-(3-h^2)X_0^2+2(1+gh)X_0Y_0={b^2\over(g+h)^2}.\eqno(5.3)$$
If $\a>0$, this is the equation of a hyperbola in the $Z=0$ plane, while
for $\a<0$ it represents an ellipse. In the degenerate case $\a=0$,
the set of critical points is given by a pair of straight lines.
Another limit case is $b=0$: also in this case
the critical points lie on a pair of straight lines.

The characteristic equation for small perturbations around these critical points
has eigenvalues $0$, $2X_0$ and $Y_0$.
Hence, for a given value of $b$, each point in the $Z=0$ plane satisfying (5.3)
with $X_0>0$, $Y_0>0$,
repels a 2-dimensional bunch of solutions in the full 3-dimensional phase space.
The points with $X_0>0$, $Y_0<0$ act instead as saddle points.
The presence of a vanishing eigenvalue is due of course to the fact that
there is a continuous set of critical points lying on a curve.

Integrating (2.16) when $\a\ne0$, one gets the expression for the radial function,
$$\ex^\r={\rm const.}\times\ex^{(\g_1\y-\g_2\c-b\x)/\a}.\eqno(5.4)$$
Recalling that for $\x\to-\inf$, $\ex^\c\sim\ex^{X_0\x}$ and $\ex^\y\sim\ex^{Y_0\x}$,
it is easy to check that the function $\ex^{\r}$ may vanish or diverge as
$\x\to-\inf$, depending on the sign of $\r_0=\g_1Y_0-\g_2X_0-b$. In the special
case $\r_0=0$, instead, $\ex^{\r}$
goes to a constant value, signalling the presence of a horizon. Combining
$\r_0=0$ with (5.3), one obtains the only real solution $X_0=Y_0=b/(g+h)^2$.
Therefore, the \cp satisfying that condition correspond to a horizon, and all the
trajectories starting from there can describe \bb solutions. The
other critical points with positive $X_0$ and $Y_0$ describe instead either naked
singularities or spatial infinity and hence the trajectories starting from
those points cannot describe \bb solutions.
This is confirmed by the computation of the Ricci scalar $\cR$ in the limit
$\x\to-\inf$: it either vanishes or diverges, except when $X_0=Y_0$, in which
case it takes a finite value.

\subsect{Critical points at infinity}

To complete the analysis of the phase space it is necessary to investigate the
nature of the critical points on the surface at infinity. These points are
attained for a finite value of $\x$ [21], and like the critical points at finite
distance may correspond to the asymptotic region of the physical solution or to
a singularity at finite distance, depending on the value of the parameters.

The analysis of the phase space at infinity can be performed defining new \coo
$u$, $y$, and $z$ such
that the surface at infinity is obtained in the limit $u\to 0$, \ie $X\to\inf$:
$$u={1\over X},\qquad y={Y\over X},\qquad z={Z\over X}\eqno(5.5)$$
In these coordinates, eqs.\ (5.1) take the form
$$\eqalign{&\dot u=-[(1+gh)\e z^2+(1+g^2)p^2]u,\cr
&\dot y=-[(1+gh)\e z^2+(1+g^2)p^2]y+(3-h^2)z^2-(1+gh)p^2,\cr
&\dot z=-[(1+gh)\e z^2+(1+g^2)p^2-y],}\eqno(5.6)$$
where we have defined $p=P/X$ and a dot denotes $u\, d/d\x$.

The discussion of the properties of the critical points at infinity is complicated,
because it strongly depends on the values of the parameters $g$ and $h$.
The limits $\a=0$ and $g=-1/h$ are degenerate. In particular, the latter is
completely integrable and is solved in the appendix.

It is convenient to discuss separately the case of positive and negative $\l$.
\bigbreak
\subsect{A. $\l>0$}
For positive $\l$, the critical points with $u=0$ can be classified in four
categories:
\medskip
1) A critical point, which we denote $L$, is placed at $y_0=-{1+gh\over\ 1+g^2}$,
$z_0^2=0$, with $p_0^2={1\over1+g^2}$. The eigenvalues of the system obtained
linearizing (5.6) around this point are $-1$, $-1$, $-{\g_1\over1+g^2}$.
\smallskip
The point $L$ is the endpoint of the trajectories lying in the $Z=0$ plane.
The analysis of stability shows that it acts
as an attractor on the trajectories coming from finite
distance and, if $\g_1>0$, also  on a two-dimensional bunch of trajectories
lying on the surface at infinity (otherwise as a saddle point).
\medskip
2) If $h^2<3$, a critical point $M$ lies at $y_0={\ 3-h^2\over1+gh}$,
$z_0^2={3-h^2\over(1+gh)^2}$, with $p_0^2=0$. The eigenvalues of the linearized
system are $-{\ 3-h^2\over1+gh}$ (double), $-{2\g_2\over1+gh}$.
\smallskip
This is the endpoint of the trajectories lying on the hyperboloid $P=0$.
The analysis of stability shows that, if $1+gh>0$, it attracts  the
trajectories coming from finite distance and a one- or a two-dimensional bunch of
those lying on the surface at infinity, depending on the sign of $\g_2$.
\medskip

3) If  $\a$, $\g_1$ and $\g_2$ all have the same sign, a critical point $N$
lies at $y_0=1$, $z_0^2={\g_1\over\a}$, with $p_0^2={\g_2\over\a}$. The eigenvalues
of the linearized system are $-1$, $-\ha(1\pm\sqrt{1+8\g_1\g_2/\a})$.

\smallskip
The point $N$ is the endpoint of the hyperboloid (5.2) in the $X=Y$ plane.
All the eigenvalues containing are real and if $\a,\g_1,\g_2<0$ are negative,
and $N$ acts as an attractor both on the trajectories coming from finite distance
and on the trajectories at infinity. In the other case, it acts as a saddle point
for the trajectories at infinity.
\medskip

4) If $\a>0$, two critical points $Q_{1,2}$ lie at $y_0=-{1+gh\pm\sqrt\a\over1+g^2}$,
$z_0^2=0$, with $p_0^2=0$. The eigenvalues of the linearized system are 0, 2, $y_0$.
If $h^2\ge3$, the critical values of $y$ both have the same sign (negative if
$1+gh>0$), otherwise have opposite sign.

\smallskip
The points $Q$ are the endpoints of the trajectories with $P=Z=0$.
They act as centers on the trajectories coming from finite distance,
while their nature for the trajectories at infinity depends on the
sign of $3-h^2$ and $1+gh$.

\medskip
\subsect{B. $\l<0$}
For negative $\l$, the critical points at $u=0$ can again be divided in four
categories:
\medskip
1) The critical point $L$ is still present, and has the same properties as for
positive $\l$.
\medskip

2) If $h^2>3$, a critical point $\overline M$ lies at $y_0={\ 3-h^2\over1+gh}$,
$z_0^2={h^2-3\over(1+gh)^2}$, with $p_0^2=0$. The eigenvalues of the linearized system
are the same as for the point $M$, but now the sign of the eigenvalues is different:
if $1+gh<0$, the point attracts the trajectories coming from finite distance and acts
either as an attractor or as a saddle point on those lying on the surface at infinity,
depending on the sign of $\g_2$.
\medskip

3) If  $\a<0$, $\g_1>0$ and $\g_2<0$, a critical point $\overline N$
lies at $y_0=1$, $z_0^2=-{\g_1\over\a}$, with $p_0^2={\g_2\over\a}$. The eigenvalues
of the linearized system are the same as for the point $N$, and all take negative
values, and hence $\bar N$ act as an attractor for all trajectories.
\medskip

4) If $\a>0$, the two critical points $Q_{1,2}$ are still present, with the same
properties as for $\l>0$.
\medskip\noindent
\bigbreak

\subsect{Asymptotic properties of the solutions}

As mentioned above, critical points at infinity correspond to the limit $\x\to\x_0$,
where $\x_0$ is a finite constant. It is easy to see that for $\x\to\x_0$, the
functions $\c$ and $\y$ behave as
$$\ex^\c\sim\xx^{-1/v_0}\qquad\qquad\ex^\y\sim\xx^{-y_0/v_0},\eqno(5.7)$$
where $v_0\id(1+gh)\e z_0^2+(1+g^2)p_0^2$.

In order to discuss the properties of the solutions, one must first of all investigate
the behavior of the radial function $\ex^\r$, that at infinity behaves as
$\xx^{-{\g_1y_0-\g_2\over\a v_0}}$. Its behavior near the \cp is reported in table 1.
It results that $\ex^\r$ diverges at points $M$ and at points $N$ if $\a>0$,
indicating the presence of an asymptotic region.
In all other cases, $\ex^\r$ vanishes at the critical points at infinity.

In order to investigate the causal structure of the solutions, it is also useful to compute
the behavior of the Ricci scalar $\cR$ at the critical points. It can be shown that this is
given by
$$\cR\sim|\x-\x_0|^{-2\[{(4+3g^2+gh)y_0+(h^2+3gh)\over\a v_0}+1\]},\eqno(5.8)$$
and is reported in table 1 as well.
The curvature is always regular at points $M$, and at points $N$ if $\a>0$ and $g(g+h)>0$,
otherwise it diverges.

From (5.7) and (2.16), (2.18), one can also deduce the behavior of the metric functions
ending at the critical
points. It is useful to write them in terms of the \sch coordinate $r$ defined
as $\int\ex^{2\n+2\r}d\x$ in (3.8), whose \ab for $\x\to\x_0$ is given in table 1 and is
analogous to that of $\ex^\r$.
Straightforward algebraic manipulations then lead to the asymptotic behavior of the
metric functions listed in table 2. The parameter $\b$ is defined in (3.23).

\bigskip
\centerline{\vbox{\offinterlineskip\vskip3pt
\halign{$\quad#\qquad$\strut&\vrule$\qquad#$\hfil&$\qquad#$\hfil&$\qquad#$\hfil&$\qquad#$
\hfil\cr&\hfil\ex^\r&\hfil\cR&\hfil r\cr
\noalign{\hrule}
L&\xx^{1/(1+g^2)}&\xx^{-2(2+g^2)/(1+g^2)}&\xx\cr
M,\overline M&\xx^{-1/(3-h^2)}&\xx^{2h^2/(3-h^2)}&\xx^{-(1+h^2)/(3-h^2)}\cr
N,\overline N&\xx^{-(g+h)^2/\a}&\xx^{4h(g+h)/\a}&\xx^{-\b/\a}\cr}}}
\bigskip
{\small Table 1: The asymptotic behavior of $\ss \ex^\r$, $\ss\cR$ and $\ss r$ for $\ss\x\to\x_0$.}
\bigskip\bigskip
\centerline{\vbox{\offinterlineskip\vskip3pt
\halign{$\quad#\qquad$\strut&\vrule$\qquad#$\hfil&$\qquad#$\hfil&$\qquad#$\hfil&$\qquad#$
\hfil\cr&\hfil\er&\hfil\en&\hfil\ef\cr
\noalign{\hrule}
L&r^{2/(1+g^2)}&r^{-2/(1+g^2)}&r^{-2g/(1+g^2)}\cr
M,\overline M&r^{2/(1+h^2)}&r^{2/(1+h^2)}&r^{2h/(1+h^2)}\cr
N,\overline N&r^{2(g+h)^2/\b}&r^{2(4+g^2-h^2)/\b}&r^{4(g+h)/\b}\cr}}}
\bigskip
{\small Table 2: The asymptotic behavior of $\ss\er$, $\ss\en$ and $\ss\ef$ as functions of
$\ss r$ for $\ss\x\to\x_0$.}
\bigskip

Although the critical points at infinity do not always correspond to asymptotic regions
of the solutions, they always give the \ab of the background solutions, \ie of those
solutions that do not depend on free parameters. For the general solutions instead, as
seen in sect.\ 3,  the mass term can dominate at infinity on the background term,
modifying the asymptotic behavior.

In fact,
the behaviors of the generic solutions in table 2 coincide with those of the exact
\bg solutions found in sect.\ 3: more precisely,
point $M$ corresponds to case $A$, point $L$ to case $B$, and point $N$
to case $C$.

\medskip
From the results listed above, we can deduce the global properties of the solutions
in terms of the values of the parameters $\l$, $g$ and $h$.
The following picture of the phase space emerges: solutions with regular horizons are
described by trajectories that
connect the point of the hyperbola (or ellipse if $\a<0$) of critical points at finite
distance such that $X_0=Y_0=b/(g+h)^2$, $b>0$ with one of the critical points at infinity.
Among these trajectories, only those that end at \cp at infinity for which $\ex^\r\to\inf$,
$\cR\to0$ correspond to regular black branes. It follows that
regular \bb solutions exist only if $\l>0$, with \ab of type $M$, if $h^2<3$, or $N$ if
$\a,\g_1,\g_2>0$ and $g(g+h)>0$.

Under these conditions the \cp at infinity attract either a 1-dimensional or a
2-dimensional bunch of trajectories in phase space. Each solution is associated with the
two free parameters $b$ (related to the mass) and $Q$ (electric charge).
In general, a third parameter  may
be necessary to parametrize the solutions. However, as emerges
from the discussion of the exact special solution in the appendix, consistency of the
thermodynamical interpretation may require that this parameter be related to the
electric charge. Also interesting is the possibility that solutions presenting different
asymptotic behaviors exist for given values of the parameters $\l$, $g$ and $h$.

\section{6. Final remarks}
Although in the general case it is not possible to obtain the planar solutions of
the EML model in analytic form, we have discussed their global properties and classified
all the possible regular \bb solutions in four dimensions.
Although several possibilities may arise, depending
on the values of the parameters $\l$, $g$ and $h$ that define the model, we have been able
to show that regular \bb solutions can exist only for a very limited range of parameters.
In particular, no regular solution exists if $\l\le0$.
Moreover, only two kinds of \ab allowed: one of them is common with the limit
of vanishing charge, while the other is characteristic of the general case. For given
values of the parameters, solutions presenting both asymptotic behaviors can exist.
No \as flat, \ads or Lifshits
solutions arise, except in the trivial case $h=0$.

The general analytic solution can be found for vanishing charge or potential, and for
special values of the parameters $g$ and $h$. In addition, also some special exact
solutions can be found for generic values of the parameters.
Some of these solutions had already been obtained in the literature [2], but not
in full generality and in awkward coordinates.

\section{Acknowledgements}
I wish to thank Mariano Cadoni for some useful discussions.

\section{APPENDIX}
When $g=-h^\mo$, the system can be completely integrated. This has been partially
done in [2], but with a choice of coordinates that obscures the structure of the
solutions.
Our discussion illustrates the possibility that for a given value of $h$ solutions
exhibiting different asymptotic behaviors exist, as it has been deduced in the
general case from the study of the phase space.

For $g=-h^\mo$, the system (2.5)-(2.6) diagonalizes,
$$\y''=(3-h^2)\,\ee,\qquad \c''=(1+h^{-2})\,\ec,\eqno(A.1)$$
and is solved by
$$\l\ee={4a^2\eax\over(3-h^2)(1-\eax)^2},\qquad Q^2\ec={4b^2K^2\ebx\over(1+h^{-2})
(1-K^2\ebx)^2},\eqno(A.2)$$
and
$$\r'={\y'\over3-h^2}-{\c'\over1+h^{-2}}+c,\eqno(A.3)$$
with $a$, $b$, $c$ and $K$ integration constants.
Substituting in the constraint (2.12), and requiring the existence of a regular horizon,
one gets
$$a=b=-{(3-h^2)(1+h^2)\over(1-h^2)^2}\ c.\eqno(A.4)$$
One must now distinguish the generic solutions with $K\ne1$ from the special solutions
with $K=1$.

Defining $r=\int\ex^{2\n+2\r}d\x\approx(1-\eax)^{-(1+h^2)/(1-h^2)}$, the $K=1$ solution
can be put in the \sch form (3.6) with
$$U={(1+h^2)^2\l\ \over(3-h^2)}\ r^{2(1+4h^2-h^4)/(1+h^2)^2}\(1-{\m\over r^\d}\),$$
$$R^2=\sqrt{(1+h^2)Q^2\over h^2(3-h^2)\l}\ r^{2(1-h^2)^2/(1+h^2)^2},
\qquad\ef=r^{-4h(1-h^2)/(1+h^2)^2},\eqno(A.5)$$
where $\d=(3-h^2)/(1+h^2)$ and $\m$ is a free parameter. These solutions exist for $h^2<3$
if $\l$ is positive and for $h^2>3$ if $\l$ is negative.
In the first case, they represent a \bb with domain wall asymptotics, parametrized by
the mass density $m={1\over h}\sqrt{{(1+h^2)^5\over(3-h^2)^3}\l Q^2}\ \m$ and the charge
density $Q$. In the second case, a naked singularity is present for $r\to\inf$.

If $K\ne1$, proceeding in the same way, one obtains
$$\eqalign{U&={(1+h^2)^2\l\ \over(3-h^2)}\ r^{2/(1+h^2)}\(1-{\m\over r^\d}\)
\(1+{\n\over r^\d}\)^{-2h^2/(1+h^2)},\cr
R^2&=\sqrt{(1+h^2)Q^2\over h^2(3-h^2)\l}\ {r^{2/(1+h^2)}\over\sqrt{\n(\n+\m)}}
\(1+{\n\over r^\d}\)^{2h^2/(1+h^2)},\cr
\ef&=r^{2h/(1+h^2)}\(1+{\n\over r^\d}\)^{2h/(1+h^2)}.}\eqno(A.6)$$
These solutions present a further free parameter $\n\ne0$, that can be associated to the
scalar charge. For $\m,\n>0$, they describe a 3-parameter family of black branes with domain
wall \ab different from that of the $K=1$ solutions, which are
recovered in the singular limit $\n\to\inf$.
The curvature is finite at the point $r=\m^{1/\d}$, that can be identified with a horizon,
while it diverges at $r=0$ and $r=-\n$, which are singularities of the metric.

The interpretation of this solution is however not easy, in particular for what concerns its
thermodynamics, because of the factor $1/\sqrt{\n(\n+\m)}$ in $R^2$, that implies that there is
no ground state corresponding to $\n=0$.
A similar situation occurs for the neutral black branes discussed in [14].
In our case, the problem can be solved imposing that $\n(\n+\m)=\s Q^2$, with $\s$ an arbitrary
normalization, thus reducing the number of free parameters.
While for \ssy solutions this condition is dictated by the normalization of
the volume element, in the planar case it must be imposed by hand. Notice that this choice
implies that $\m$ and $\n$ must be positive, in order to avoid naked singularities.

The special cases $h^2=1,\,3$ should
be studied separately, since for these values of the parameter $h$ some degeneracies appear in
our calculations, but the generalization is straightforward, and we do not consider it in detail.


\beginref
\ref [1] S.J. Poletti and D.L. Wiltshire, \PR{D50}, 7260 (1994).
\ref [2] C. Charmousis, B. Gout\'eraux and J. Soda, \PR{D80}, 024028 (2009).
\ref [3] C. Charmousis, B. Gout\'eraux, B.S. Kim, E. Kiritsis and R. Meyer,
\JHEP{1011}, 151 (2010).
\ref [4] E. Perlmutter, \JHEP{1102}, 013 (2011).
\ref [5] B. Gouteraux and E. Kiritsis, \JHEP{1112}, 036 (2011).
\ref [6] M. Cveti\v c, H. L\"u and C.N. Pope, \PR{D63}, 086004 (2001).
\ref [7] X. Dong, S. Harrison, S. Kachru, G. Torroba and H. Wang, \JHEP{1206},
041 (2012);
L. Huijse, S. Sachdev and B. Swingle, \PR{B85}, 035121 (2012).
\ref [8] J.D. Bekenstein, \PR{D5}, 1239 (1972); \PR{D5}, 2403 (1972).
\ref [9] S. Kachru, X. Liu and M. Mulligan, \PR{D78}, 106005 (2008).
\ref [10] G. Bertoldi, B. Burrington and A. Peet, \PR{D80}, 126003, (2009),
\PR{D80}, 126004, (2009);
G. Bertoldi, B. Burrington, A. Peet and I. Zadeh, \PR{D83}, 126006, (2011).
\ref [11] H.J. Boonstra, K. Skenderis and P.K. Townsend, \JHEP{9901}, 003 (1999);
I. Kanitscheider and K. Skenderis, \JHEP{0904}, 062 (2009).
\ref [12] M. Cadoni and P. Pani, \JHEP{1104}, 049 (2011);
 M. Cadoni, S. Mignemi and M. Serra, \PR{D85}, 086001 (2012).
\ref [13] M. Cadoni, S. Mignemi and M. Serra, \PR{D84}, 084046 (2011).
\ref [14] M. Cadoni and S. Mignemi, \JHEP{1206}, 056 (2012).
\ref [15] K. Goldstein, S. Kachru, S. Prakash and S. Trivedi, \JHEP{1008}, 078 (2010).
\ref [16] M. Cadoni, G. D'Appollonio and P. Pani, \JHEP{1003},100 (2010);
\ref [17] D.S. Fisher, \PRL{56}, 416 (1986).
\ref [18] E. Shaghoulian, \JHEP{1205}, 065 (2012);
N. Ogawa, T. Takayanagi, T. Tadashi and T. Ugajin, \JHEP{1201}, 125 (2012).
\ref [19] D. Garfinkle, G.T. Horowitz and A. Strominger, \PR{D43}, 3140 (1991).
\ref [20] G.W. Gibbons, K. Maeda, \NP{B 298}, 741 (1988).
\ref [21] D.L. Wiltshire, \PR{D36}, 1634 (1987); \PR{D44}, 1100 (1991);
S. Mignemi and D.L. Wiltshire, \CQG{6}, 987 (1989); \PR{D46}, 1475 (1992);
S. Mignemi, \PR{D62}, 024014 (2000); \PR{D74}, 124008 (2006);
M. Melis and S. Mignemi, \CQG{22}, 3169 (2005); \PR{D73}, 083010 (2006).
\endref

\end